%% file: paper_eprint.tex
\documentclass[twocolumn,showpacs,aps,prl,superscriptaddress]{revtex4}

\usepackage{graphicx}
\usepackage{dcolumn}
\usepackage{amsmath}
\usepackage{epsfig}

\input babarsym.tex
\input symbols.tex

\newcommand{\BABARPubYear}    {05}
\newcommand{\BABARPubNumber}  {033}

\newcommand{\SLACPubNumber} {11325}

\newcommand{\kstarpi}{\ensuremath{B^{0} \ra K^{*+} \pi^{-}}}

\newcommand{\dpi}{\ensuremath{B^{0} \ra D^{-} \pi^{+}}}

\newcommand{\BzBzbar}{\ensuremath{B^{0}\overline{B^{0}}}}

\def\figurebox#1#2#3{%
    \def\arg{#3}%
    \ifx\arg\empty
    {\hfill\vbox{\hsize#2\hrule\hbox to #2{\vrule\hfill\vbox to #1{\hsize#2\vfill}\vrule}\hrule}\hfill}%
    \else
    {\hfill\epsfbox{#3}\hfill}%
    \fi}

\begin{document}

\preprint{\babar-PUB-\BABARPubYear/\BABARPubNumber} 
\preprint{SLAC-PUB-\SLACPubNumber} 

\begin{flushleft}
\babar-PUB-\BABARPubYear/\BABARPubNumber\\
SLAC-PUB-\SLACPubNumber\\
\end{flushleft}

\title{
{\large \bf \boldmath
Measurements of Neutral $B$ Decay Branching Fractions to $\KS\pip\pim$ 
Final States and the Charge Asymmetry of \kstarpi}
}

%
\author{B.~Aubert}
\author{R.~Barate}
\author{D.~Boutigny}
\author{F.~Couderc}
\author{Y.~Karyotakis}
\author{J.~P.~Lees}
\author{V.~Poireau}
\author{V.~Tisserand}
\author{A.~Zghiche}
\affiliation{Laboratoire de Physique des Particules, F-74941 Annecy-le-Vieux, France }
\author{E.~Grauges}
\affiliation{IFAE, Universitat Autonoma de Barcelona, E-08193 Bellaterra, Barcelona, Spain }
% ...
\collaboration{The \babar\ Collaboration}
\noaffiliation

\date{\today}

\begin{abstract}
We analyze the decay $\Bz \to \KS\pip\pim$ using a sample of 232 
million $\Upsilon(4S) \rightarrow$ \BB\ decays collected with the \babar\ 
detector at the SLAC PEP-II asymmetric-energy $B$ factory. A maximum 
likelihood fit finds the following branching fractions:
${\cal B}$($B^0 \to K^0\pi^+\pi^-$) = (43.0 $\pm$ 2.3 $\pm$ 2.3) $\times$ 
10$^{-6}$, ${\cal B}(B^0 \to f_0 (\to \pip\pim)K^0) =  (5.5 
\pm 0.7 \pm 0.5 \pm 0.3) \times 10^{-6}$ and ${\cal B}$($B^0$ 
$\rightarrow$ $K^{*+}$$\pi^-$) = (11.0 $\pm$ 1.5 $\pm$ 0.5 $\pm$ 0.5) 
$\times$ 10$^{-6}$. For these results, the first uncertainty is statistical, 
the second is systematic, and the third (if present) is due to the effect of 
interference from other resonances.  We also measure the \CP-violating charge asymmetry
in the decay $B^0$ $\rightarrow$ $K^{*+}$$\pi^-$, ${\cal A}_{K^*\pi} =  -0.11 \pm 0.14 \pm 0.05$. 
\end{abstract}

\pacs{13.25.Hw, 12.15.Hh, 11.30.Er}

\maketitle

Measurements of charmless three-body $B$ decays, which are dominated by their 
intermediate quasi-two body decays, are important in furthering our
understanding of quark couplings described by the 
Cabibbo-Kobayashi-Maskawa matrix~\cite{ckm}.
\CP\ violation can be probed through the investigation of
neutral $B$-meson decays to resonance channels with the final state $\Ks\pi^+\pi^-$,
such as $f_0\Ks$~\cite{f0Ks}, $\rho^0\Ks$~\cite{rho0Ks} and $K^{*+}\pi^-$~\cite{kspipi}. 

By measuring the charmless branching fraction of $B^0 \to \Ks\pi^+\pi^-$,
along with those of its dominant resonant sub-modes, we can 
obtain information about the structure of the decay Dalitz plot.  
Such measurements have previously been performed by the CLEO~\cite{cleo}, Belle~\cite{belle} 
and \babar~\cite{kspipi,f0Ks,rho0Ks} experiments. 

QCD factorization models~\cite{beneke} have predicted branching
fractions and asymmetries for charmless $B$ decays.
Predictions have also been made using flavor
SU(3) symmetry~\cite{isospin}.  For 
$B^0 \rightarrow K^{*+}\pi^-$, predictions~\cite{guo} have been 
made for the branching fractions and charge asymmetry,
\begin{equation}
{\cal A}_{K^*\pi} = \frac{\Gamma_{\Bzb\to K^{*-}\pi^+} - \Gamma_{\Bz
    \to K^{*+}\pi^-}}{\Gamma_{\Bzb \to K^{*-}\pi^+} + \Gamma_{\Bz
    \to K^{*+}\pi^-}} ~,
\end{equation}
which is a \CP-violating quantity since the decay channel is a 
flavor eigenstate. \CP\ violation in charge asymmetry has 
already been observed by \babar\ and Belle in $\Bz \to K^+ \pim$~\cite{kpi}.

In this paper the branching fractions of $B^0 \to K^0\pi^+\pi^-$, $B^0$ 
$\rightarrow$ $K^{*+}$$\pi^-$ and $B^0$ $\rightarrow$ $f_0(980)(\rightarrow 
\pi^+\pi^-)K^0$ are presented, averaged over charge-conjugate states,
along with a measurement of the charge asymmetry in $B^0$ $\rightarrow$ 
$K^{*+}$$\pi^-$. The selection criteria require events with a 
reconstructed \KS in the final state.  Results are stated in terms of
the $K^0$ final state, taking into account the probabilities for 
${\cal{B}}(K^0\to\KS)$ and ${\cal{B}}(\KS\to\pip\pim)$~\cite{pdg}.  
For the $B^0 \to K^0\pi^+\pi^-$ branching fraction, the total charmless 
contribution to the Dalitz plot is measured (with charmed and charmonium
resonances removed), including contributions from resonant charmless 
sub-structure. 

The data used in this analysis were collected at the \pep2\ 
asymmetric-energy \epem\ storage ring with the \babar\ detector~\cite{babar}. The 
\babar\ detector consists of a double-sided five-layer silicon tracker, 
a 40-layer drift chamber, a Cherenkov detector, an electromagnetic calorimeter 
and a magnet with instrumented flux return. The data sample has an integrated 
luminosity of 210~\invfb\ collected at the $\FourS$ resonance, which 
corresponds to $(231.8 \pm 2.5)\times 10^6$ \BB\ pairs. It is assumed that 
the $\FourS$ decays equally to neutral and charged $B$-meson pairs. In 
addition, 21.6~\invfb\ of data collected at 40~MeV below the $\FourS$ 
resonance were used for background studies.

The reconstruction of candidate $B$ mesons combines two charged tracks
and a $\KS$ candidate, with the $\KS$ being reconstructed from
two oppositely charged tracks consistent with $\pip\pim$.  
The $\Bz$ decay vertex is reconstructed from the 
two charged tracks that were not daughters of the $\KS$, with the
requirements that the tracks originate from the beam-spot,
have at least 12 hits in the drift chamber and 
have a transverse momentum greater than 100~\mevc.  \KS\ candidates are 
required to have a reconstructed mass within 15\mevcc
of the nominal \KS\ mass~\cite{pdg}, at least a five standard deviation separation
between the \Bz\ decay vertex and its own decay vertex, and a cosine of
the angle between the line joining the \Bz\ and \KS decay vertices 
and the \KS\ momentum vector greater than 0.999.  To identify pions we
use measurements of energy loss (\dedx) in
the tracking system, the number of photons detected by the
Cherenkov detector and the corresponding Cherenkov angle.
Candidate pions must fail the electron selection, which is based
on \dedx measurements, shower shape in the calorimeter, and the ratio of
energy in the calorimeter to momentum in the drift chamber.
Using simulated Monte Carlo (MC) events, we determine an approximate
mean and width ($\sigma$) of the mass distribution for the resonances, and
choose the resonance band to be $\pm3\sigma$ from the mean.
For the decay $B^0$ $\rightarrow$ $K^{*+}$$\pi^-$ we require 
0.776 $<$ $m_{\Ks\pi}$ $<$ 1.010~GeV/$c^2$ and for 
$B^0$ $\rightarrow$ $f_0 \KS$ we require 0.879 $<$ $m_{\pip\pim}$ $<$ 1.069~GeV/$c^2$.  

The dominant source of background is continuum quark production 
($e^+e^-$ $\rightarrow$ $q\bar{q}$ where $q$ = {\em u,d,s,c}).
An event-shape variable, the cosine of the angle $\theta_T$ between 
the thrust axis of the selected $B$ candidate and the thrust axis of the 
rest of the event~\cite{babar}, is used to suppress this background. 
The distribution of $|\cos\theta_T|$ is strongly peaked towards unity for 
continuum background but is flat for signal events. The requirement
$|\cos\theta_T| < 0.9$ reduces the relative amount
of continuum background.

To separate signal events from the remaining background events, we use
two kinematic variables and one event-shape variable.  The first 
kinematic variable 
$\DeltaE$, is the difference between the center-of-mass (CM) energy of the $B$ 
candidate and $\sqrt{s}/2$, where $\sqrt{s}$ is the total CM energy of the
\epem\ beams. The second is the beam-energy-substituted 
mass $\mes = \sqrt{(s/2 + \pvec_i \cdot \pvec_B)^2/E_i^2 - \pvec^2_B}$, where
$\pvec_B$ is the $B$ momentum and  ($E_i, \pvec_i$) is the four-momentum of 
the $\FourS$ in the laboratory frame. We require these variables to be
in the ranges $|\Delta E| <0.1 \gev$ and $5.22<\mes<5.29 
\gevcc$. We construct a Fisher discriminant ($\mathcal{F}$)~\cite{Fisher} using 
a linear combination of five event-shape variables: the cosine of the angle 
between the $B$-candidate momentum and the beam axis, the cosine of the 
angle between the $B$-candidate thrust axis and the beam axis, the zeroth 
and second angular moments of the energy flow about the thrust axis of the 
$B$~\cite{f0Ks}, and the output of the $B$-flavor tagging algorithm,
which uses the information from the other $B$~\cite{rhorho}.  
This forms a more efficient Fisher discriminant than used in our previous
measurement, Ref.~\cite{kspipi}.

Other $B$-meson decays can mimic a $\KS\pip\pim$ final state.  MC
events are used to identify the $B$ decays that contribute background events
to the data sample, and we use the available information on exclusive 
measurements~\cite{pdg,hfag} to find how many events from this background 
to expect in the data set.  The largest $B$ background is seen to come from 
quasi two-body decays including charmonium mesons such as $J/\psi\Ks$, 
$\chi_{c0}\Ks$ and $\psi(2S)\Ks$.  In these cases the charmonium meson decays to 
$\pi^+\pi^-$ or to $\mu^+\mu^-$ that are misidentified as pions. 
Most of these events are removed by vetoing the reconstructed 
$\pi^+\pi^-$ masses consistent with 3.04 $<$ $m_{\pi^+\pi^-}$ $<$ 
3.16~GeV/$c^2$, 3.32 $<$ $m_{\pi^+\pi^-}$ $<$ 3.51~GeV/$c^2$ and 3.63 $<$ 
$m_{\pi^+\pi^-}$ $<$ 3.74~GeV/$c^2$, identifying the $J/\psi$, $\chi_{c0}$ 
and  $\psi(2S)$ mesons respectively.  From simulated data we estimate that 126
$\pm$ 8 $\Bz \to J/\psi \KS$ events and 6 $\pm$ 3 $\Bz \to \psi(2S)\KS$
events fall outside these vetoes, and these are included in the model.
We veto events that are consistent with 
$B^0\rightarrow D^-(\rightarrow \Ks\pi^-)\pi^+$ by excluding those with 
$1.8< m_{\KS\pi} < 1.91$ GeV/$c^2$.  However, Monte Carlo simulation shows that 
71 $\pm$ 8 $B^0$ $\rightarrow$ $D^-(\rightarrow \Ks\pi^-)$$\pi^+$ background 
events still remain, where the reconstructed $D^\mp$ mass falls outside
the veto as a result of using a $\Ks$ or a $\pi$ from the other $B$ decay in the event. 
Other incorrectly reconstructed charmed decays $B \to D^{(*)}X$ are 
also included in the model.

After the above selection criteria are applied, 12.4\% of events 
have more than one candidate that satisfies the selection criteria.  
In a signal MC study, selecting the candidate whose $\cos\theta_T$ value 
is closest to zero is found to select the true signal candidate in 69.2\% 
of such events.  These requirements result in a final sample size of 
approximately 80,000 events.

After all requirements, the largest charmless $B$ background to the 
$B^0 \to \KS\pi^+\pi^-$ measurement is the decay $B^0$ $\rightarrow$ 
$\eta^\prime\KS, \eta^\prime\to\rho^0(770)\gamma,\rho^0\to\pip\pim$,
which tends to peak in the signal region and which contributes 54
$\pm$ 19 events.  Table~\ref{tab:bbgs} shows the $B$-background
modes for the $\Bz \to K^{*\pm}\pi^\mp$ and $\Bz \to f_0\KS$ channels.
These events are effectively subtracted from the measured signal.  To measure 
the nonresonant $\Bz \to\Ks\pi^+\pi^-$, we select a region of the Dalitz 
plot believed to be free of resonances, ($3 < m_{\pip\pim} < 4 \gevcc$ 
and $ m_{\KS \pi^{\pm}} > 1.91 \gevcc$).  Backgrounds from other $B$ 
decays and from continuum events are subtracted.  Assuming a uniform 
nonresonant distribution in the Dalitz plane, we set an upper limit 
of 2.1 $\times 10^{-6}$ at a 90\% confidence level on the nonresonant 
$\Bz \to \KS\pip\pim$ branching fraction. All other branching fractions are 
taken from Refs.~\cite{pdg, hfag}.
\begin{table}[!h]
\caption{The $B$-background modes for the channels $\Bz \to K^{*\pm}\pi^\mp$ 
and $\Bz \to f_0\KS$. $\Bz \to \rho^0\KS$ is included at a level consistent 
with Ref.~\cite{rho0Ks}.  $K^{**}$ refers to heavier $K^{*}$ resonances, e.g.
$K_0^{*}(1430)$.
}\label{tab:bbgs}
\begin{tabular}{lcc}
\hline
$B$-background &
Number Expected &
Number Expected \\
Mode &
($\Bz \to K^{*\pm}\pi^\mp$) &
($\Bz \to f^0\KS$) \\
\hline
$B^0 \to K^{*\pm}\pi^\mp$ &
- &
5 $\pm$ 1 \\
$B^0$ $\rightarrow$ $f_0\KS$ &
4 $\pm$ 1 &
- \\
$\Bz \to \rho^0\KS$ & 
5 $\pm$ 2 &
14 $\pm$ 4  \\
$B^0$ $\rightarrow$ $K^{**+}\pi^-$ &
23 $\pm$ 3 &
4 $\pm$ 1 \\
Nonresonant &
7 $\pm$ 1 &
5 $\pm$ 1 \\
$\Bz \to D^{\mp}\pi^{\pm}$ &
16 $\pm$ 2 &
0 \\
$\Bz \to \eta^\prime\Ks$ &
1 $\pm$ 1 &
19 $\pm$ 7 \\
$\Bz \to \jpsi\KS$ &
6 $\pm$ 1 & 
0 \\
\hline
\end{tabular}
\end{table}

We use an extended maximum likelihood fit to extract the signal yield 
for each of the channels being investigated.
The likelihood function for $N$ events is:
\begin{equation}
{\cal L}= {\exp{(-\sum_j N_j)}}\prod_i^N ~ \left(\sum^M_{j=1} N_j P_j(\vec{x}_i)\right)
\end{equation}
\noindent where $i$ and $j$ are integers,  $M$ is the number of hypotheses 
(signal, continuum background and $B$ background), $N_j$ is the
number of events for the $j$th hypothesis determined by maximizing
the likelihood function, and
$P_{j}(\vec{x}_i)$ is a probability density function (PDF) 
evaluated using the vector $\vec{x}_i$, in this case \mes, 
\DE, and $\mathcal{F}$.  Correlations between these variables are small for 
signal and continuum background hypotheses and the total PDF is a product 
$P_{j}(\vec{x}_i) = P_{j}(\mes) \cdot P_{j}(\DE ) \cdot P_{j}(\mathcal{F})$. 
However for $B$ background, it is necessary to account for correlations 
observed between \mes\ and \DE\ by using a two-dimensional PDF for these 
variables. 

The parameters of the signal and $B$-background PDFs are determined 
from MC simulation and fixed in the fit, along with the $B$-background
normalization. The continuum background 
parameters are allowed to vary in the fit, to help reduce systematic 
effects from this dominant event type. Sideband data (which lie in 
the region $0.1 < \Delta E <0.3 \gev$ and $5.22<\mes<5.29 \gevcc$) are used to 
model the continuum background PDFs. For the \mes\ PDFs, a Gaussian 
distribution is used for signal and a threshold function~\cite{argus}
for continuum. For the \DE\ PDFs, a sum of two Gaussian distributions 
with the same means is used for the signal and a first-order polynomial
for the continuum background. Finally, for the $\cal{F}$ PDFs, a sum of 
two Gaussian distributions with distinct means and widths is used for signal 
and a sum of two Gaussian distributions with the same means is used to model
the continuum background. The Fisher discriminant distribution of the $B$ backgrounds 
is modeled by an asymmetric Gaussian distribution that has different widths above
and below the modal value.
We use $B^0$ $\rightarrow$ $D^-(\rightarrow \Ks\pi^-)\pi^+$ as a calibration
mode since it exhibits a one-to-one signal to continuum background ratio, 
allowing the
signal parameters in a fit to be floated.  A fit to these data is used in order
to quantify any corrections and uncertainties due to MC.  These
corrections are applied to the fits to the charmless data sample.

To extract the branching fractions for the decay modes $B^0$ $\rightarrow$ 
$K^{*+}$$\pi^-$ and $B^0$ $\rightarrow$ $f_0 K^0$ we use the relation
\begin{equation}
\label{eq:bf}
 \mbox{$\mathcal{B}$} \,=\ \frac{N_{sig}}{2N_{\BzBzbar} \times \varepsilon}~,
\end{equation}
\noindent where $N_{sig}$ is the number of signal events fitted, $\varepsilon$ 
is the signal efficiency obtained from MC and $N_{\BzBzbar}$ is the total 
number of \BzBzbar\ pairs. 

For the charmless $B^0 \to K^0\pi^+\pi^-$ branching 
fraction (and also for the nonresonant upper limit in the $B$-background
studies above), it is necessary to account for the variation in
efficiency, between approximately 5\% and 40\%, across the Dalitz plot
and to know how the signal events are distributed across the Dalitz plot.
To do this we assign to the $j$th event ${\cal W}_j$ = 
$\sum_iV_{sig,i}P_{i}(\vec{x}_j)/\sum_k N_{k}{P_{k}(\vec{x}_j)}$ 
where $V_{sig,i}$ are the signal components of the covariance matrix 
obtained from the fit. 
This procedure projects out the signal distributions~\cite{splot} shown
in Figures 1-4.
The branching fraction is then calculated as ${\cal B} = 
\sum_{j}{\cal W}_j/(\varepsilon_{j} \times N_{\BzBzbar})$, where 
$\varepsilon_j$ is the efficiency, as a function of Dalitz plot position,
simulated in small bins using high statistics MC.
\begin{figure}[!h]
\resizebox{\columnwidth}{!}{
\begin{tabular}{cc}
\includegraphics{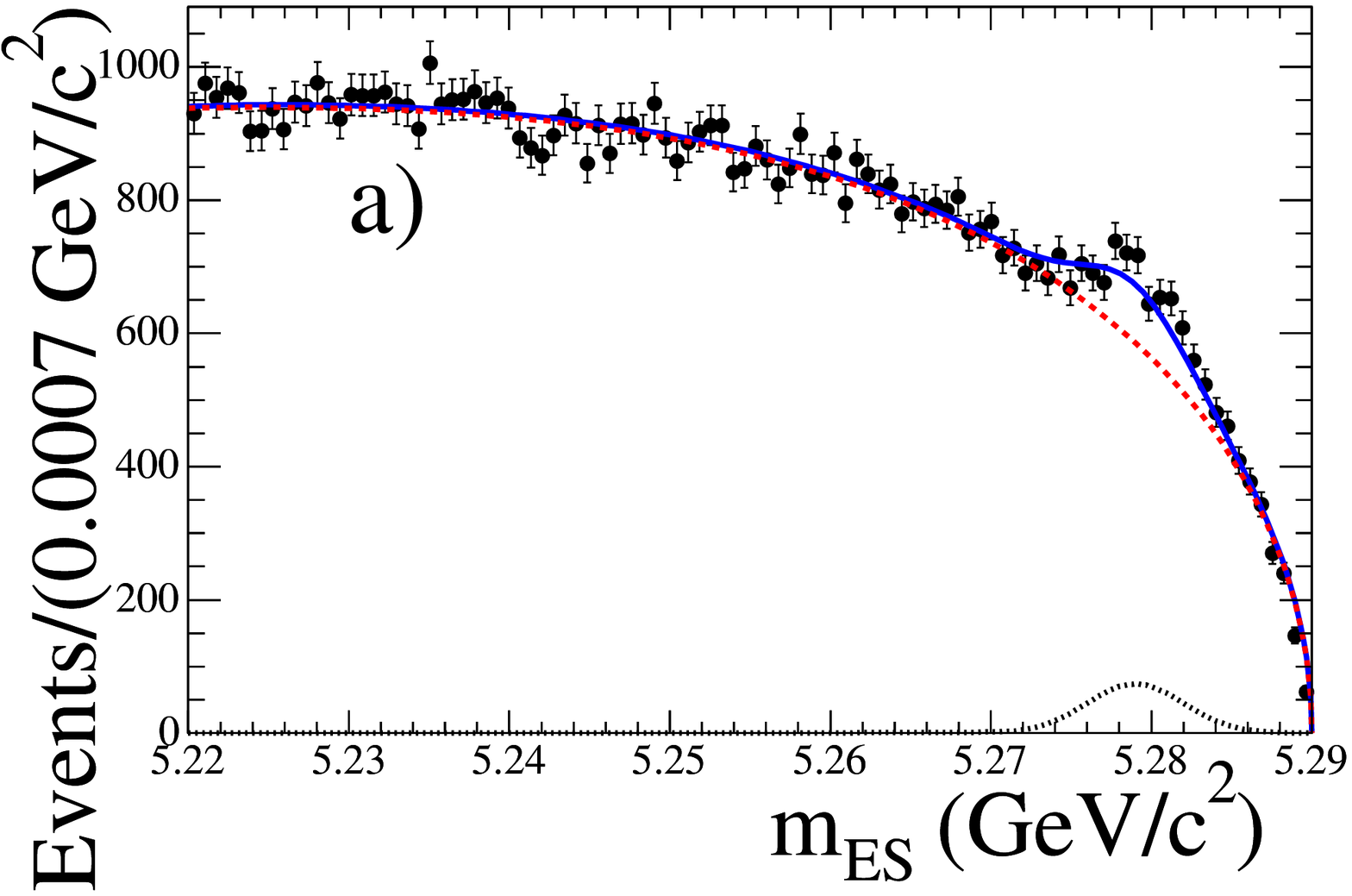} &
\includegraphics{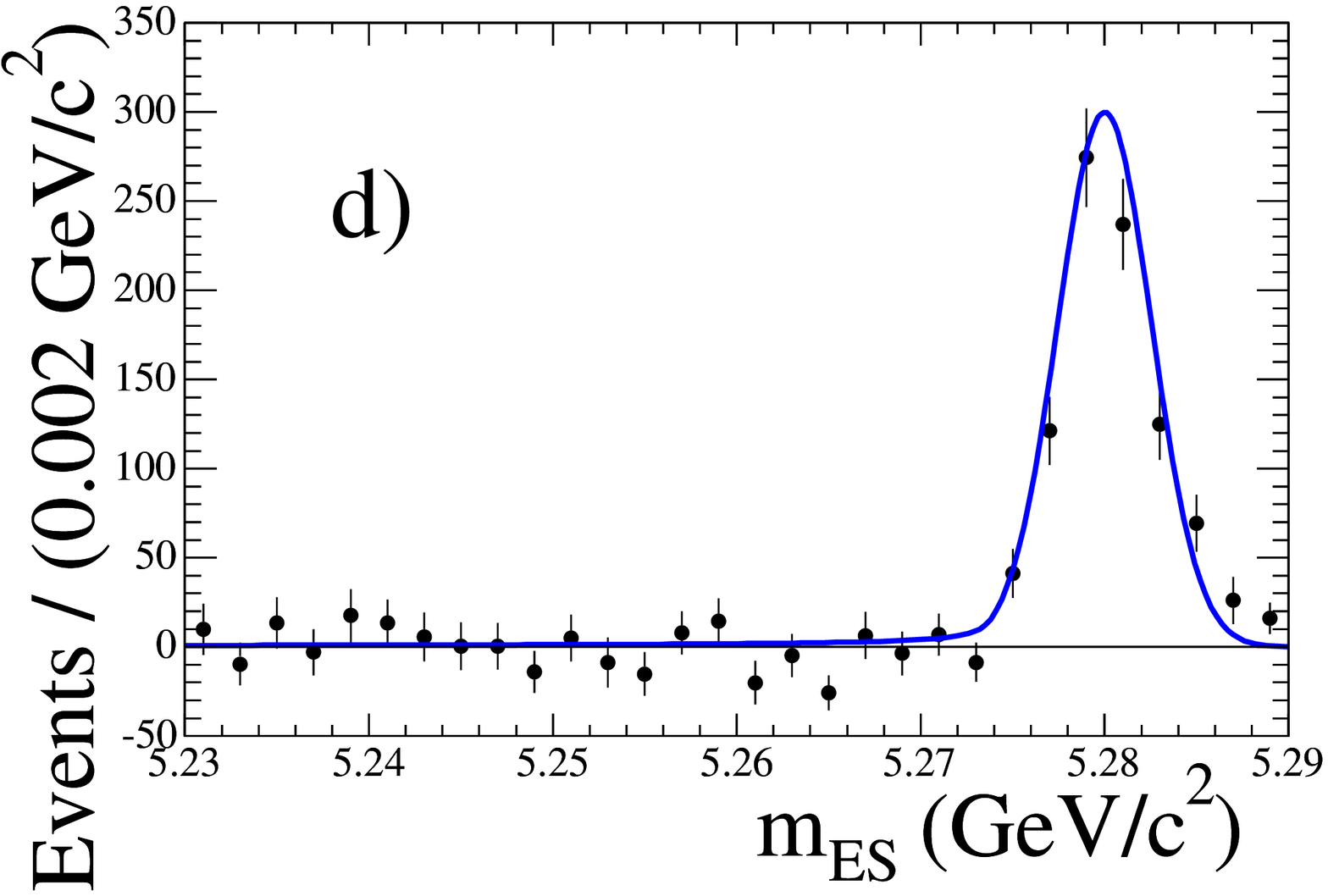} \\
\includegraphics{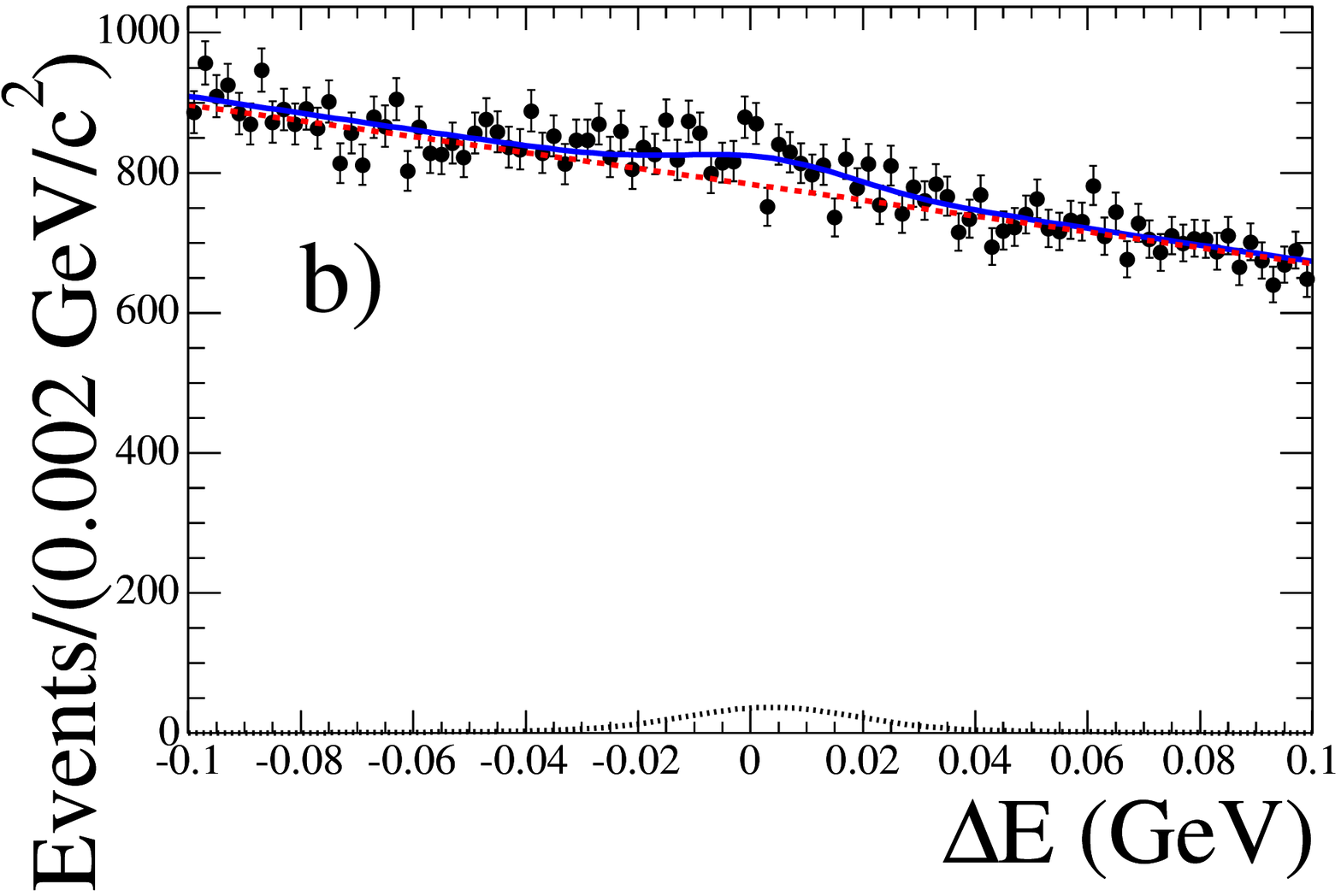}&
\includegraphics{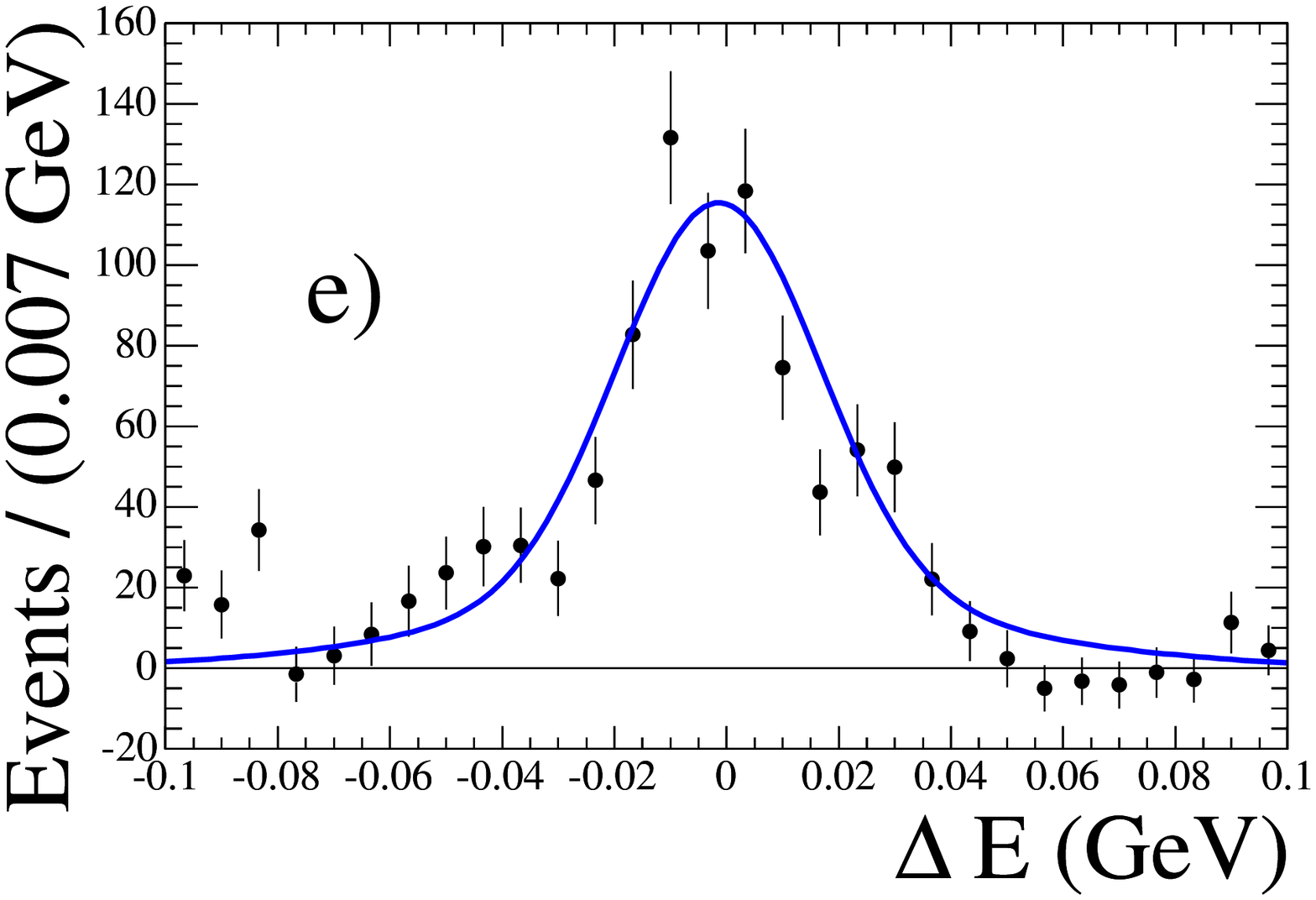} \\
\includegraphics{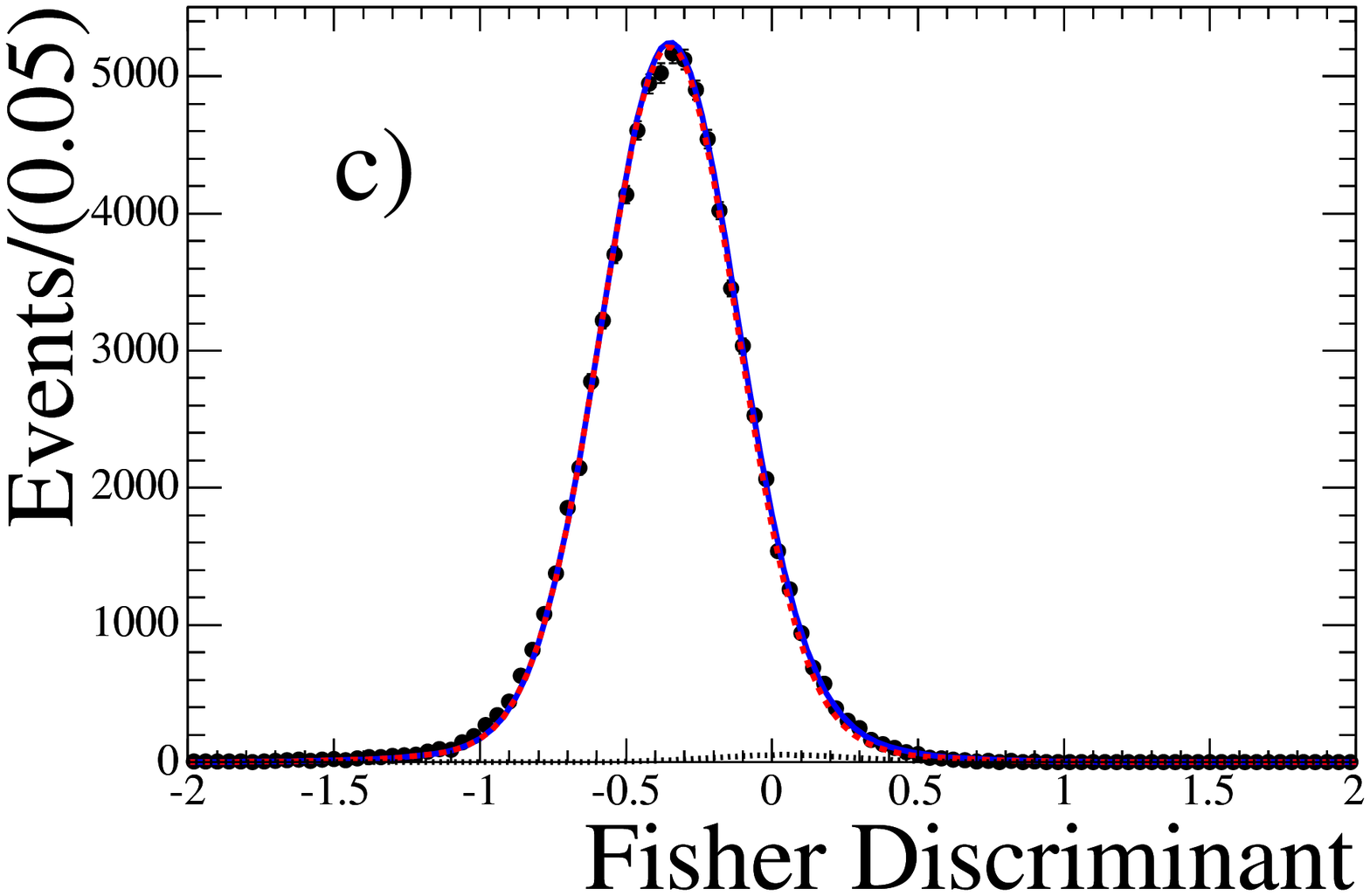}&
\includegraphics{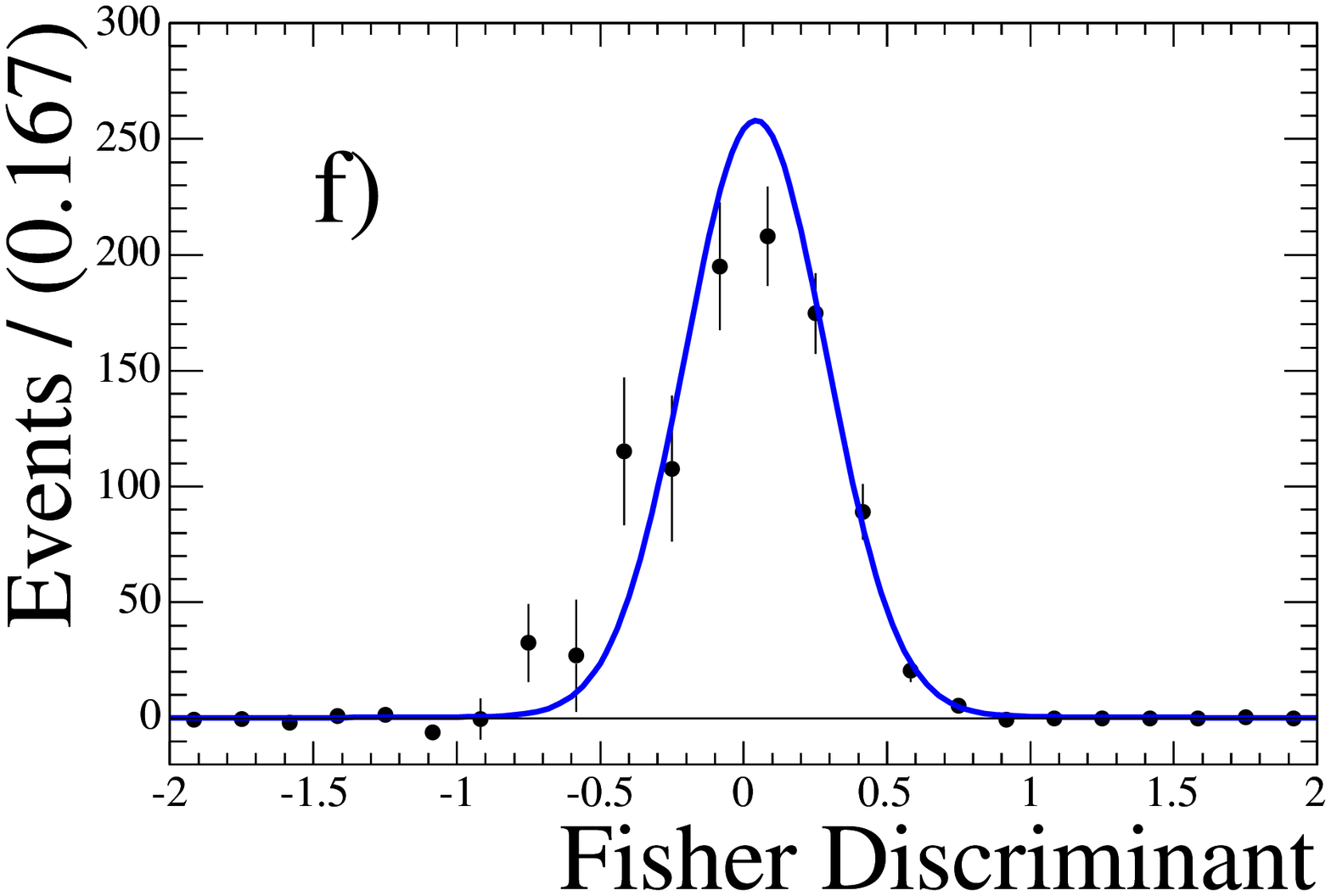} \\
\end{tabular}
}
\caption{
Plots of the maximum likelihood fit to data 
for $\Bz \to K^0 \pip\pim$ candidates.  Plots a)-c) show
the distributions of all events that pass the selection criteria 
for (a) \mes\, (b) \de\, and (c) Fisher, with the solid (blue) 
line indicating the total model, the (red) dotted line indicating 
shape of the continuum background model and the (black) dashed line 
indicating the signal model.  Plots d)-f) show the signal 
distributions for (d) $m_{ES}$, (e) \DE\, and  (f) Fisher,  
where the (black) circles are the signal distribution~\cite{splot} and
the solid (blue) curve is the signal PDF that was fitted in the maximum likelihood fit.  
}\label{fig:fitproj}
\end{figure}
\begin{figure}[!h]
\resizebox{\columnwidth}{!}{
\begin{tabular}{cc}
\includegraphics{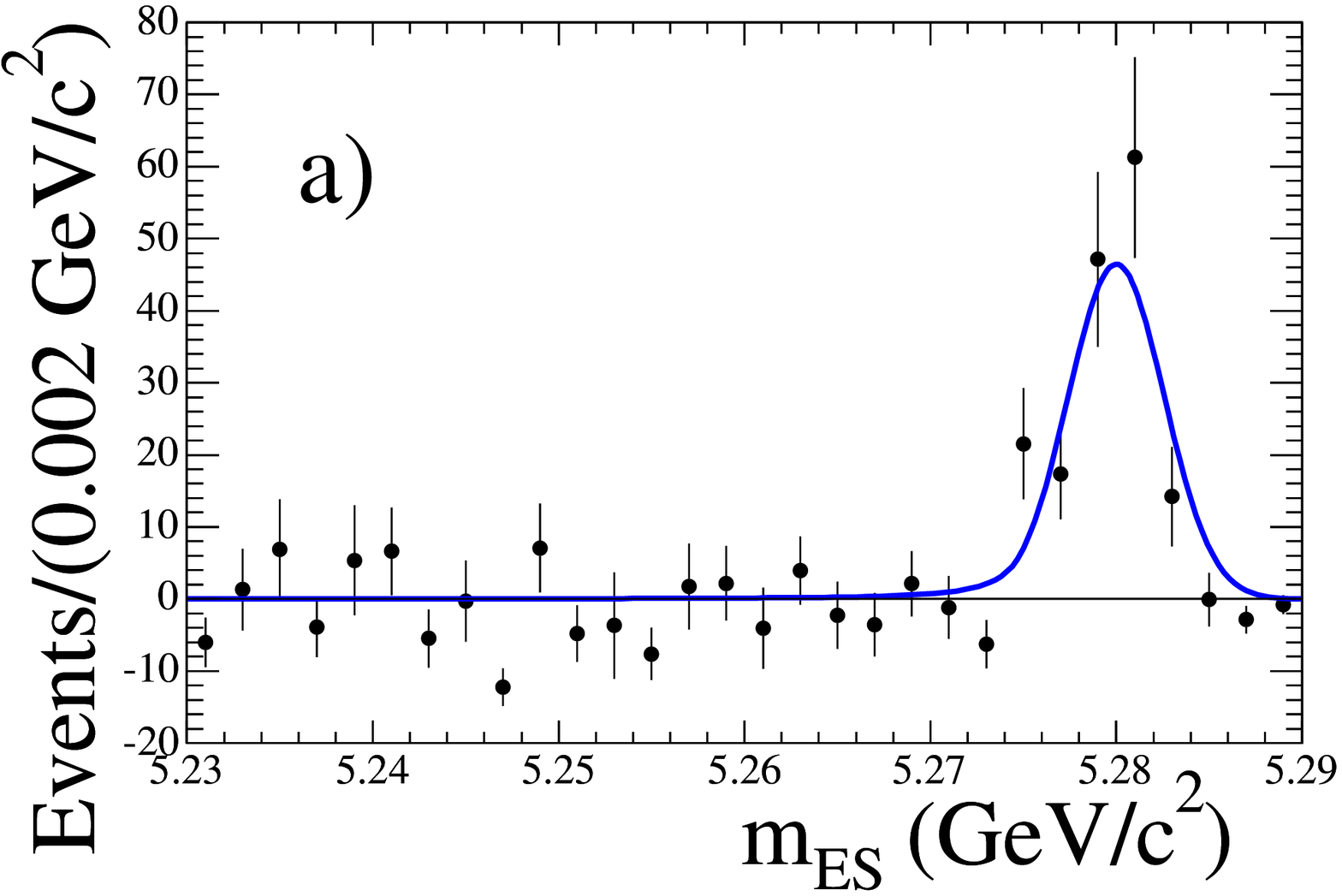} & 
\includegraphics{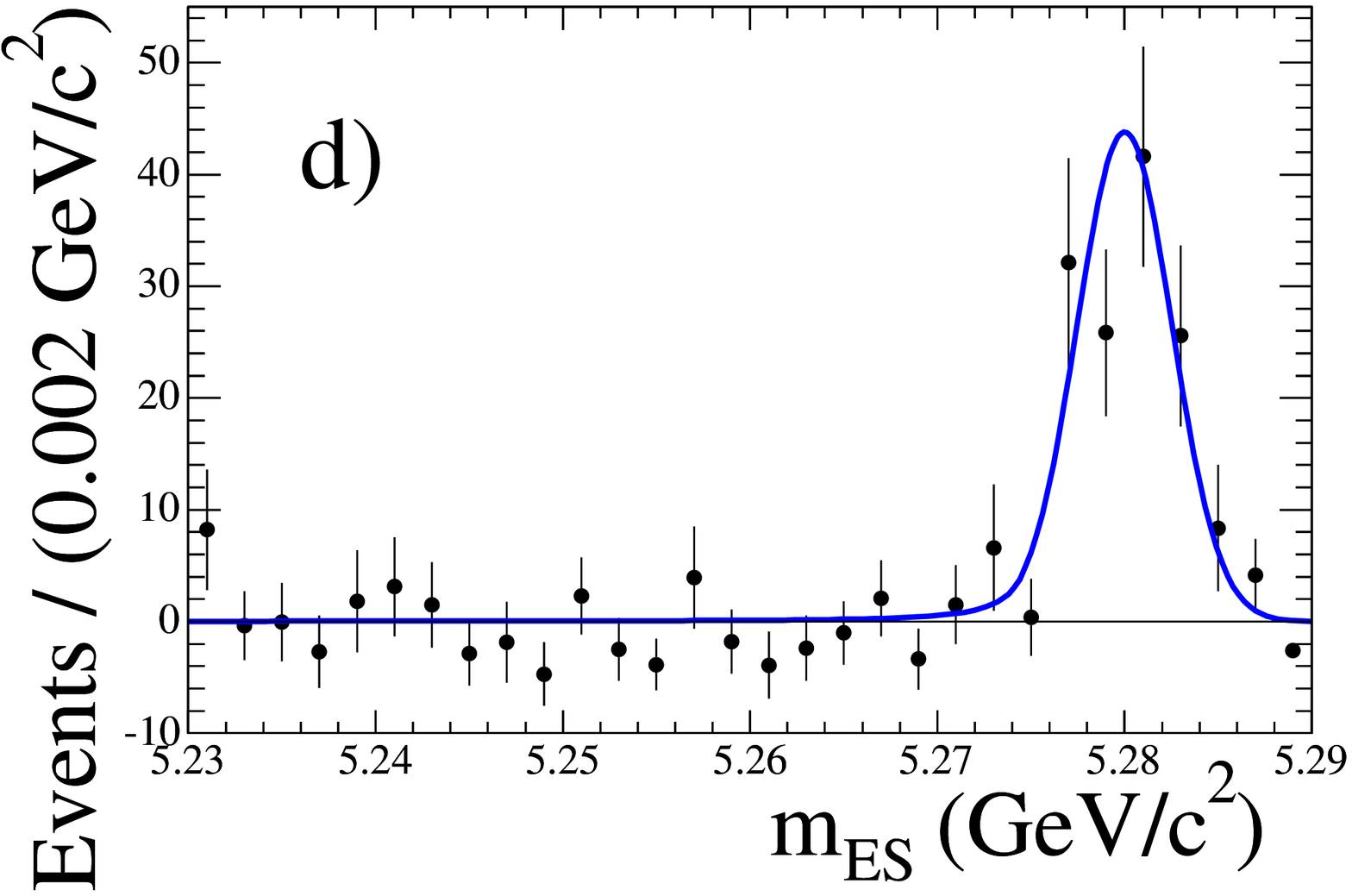} \\  
\includegraphics{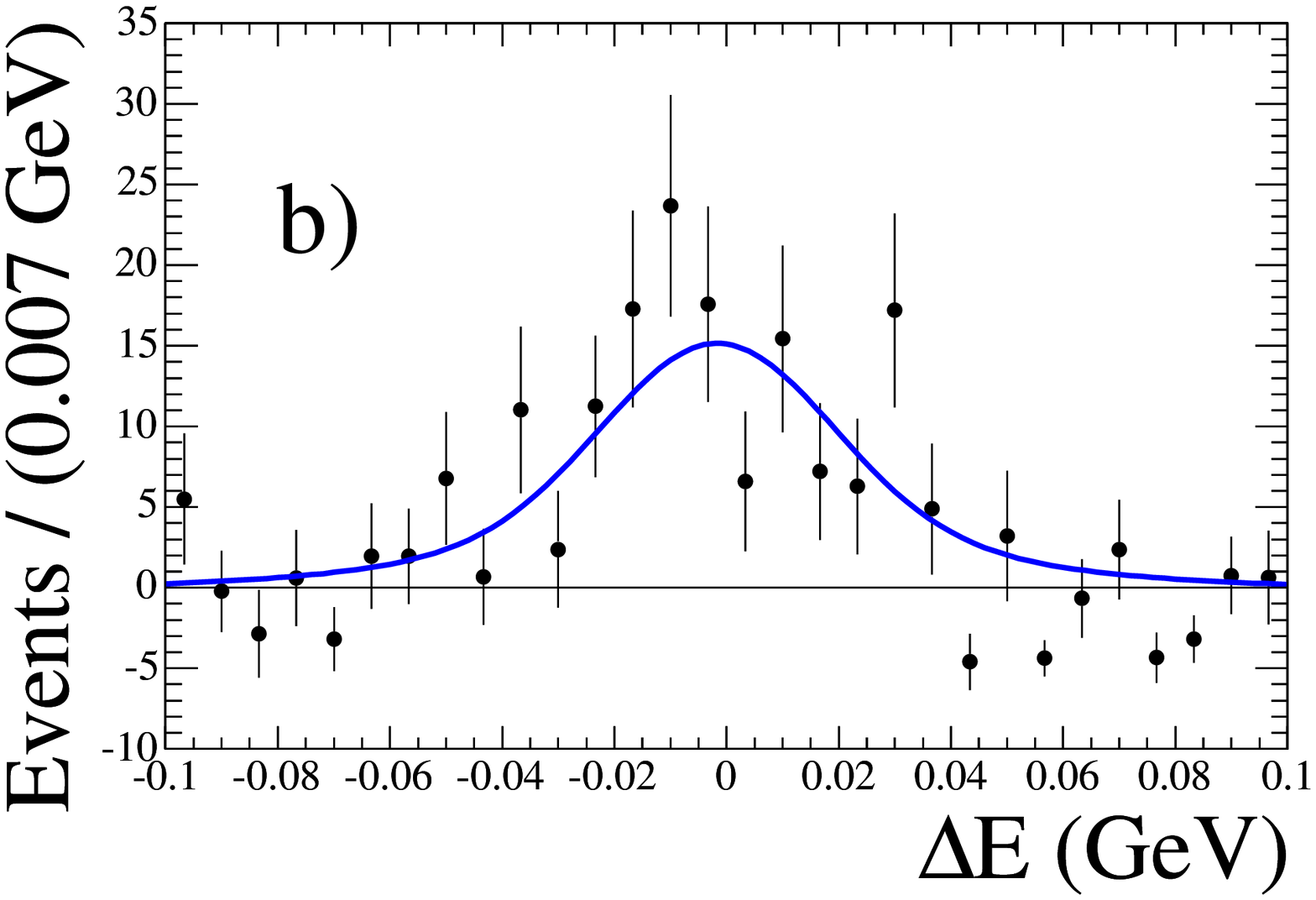} & 
\includegraphics{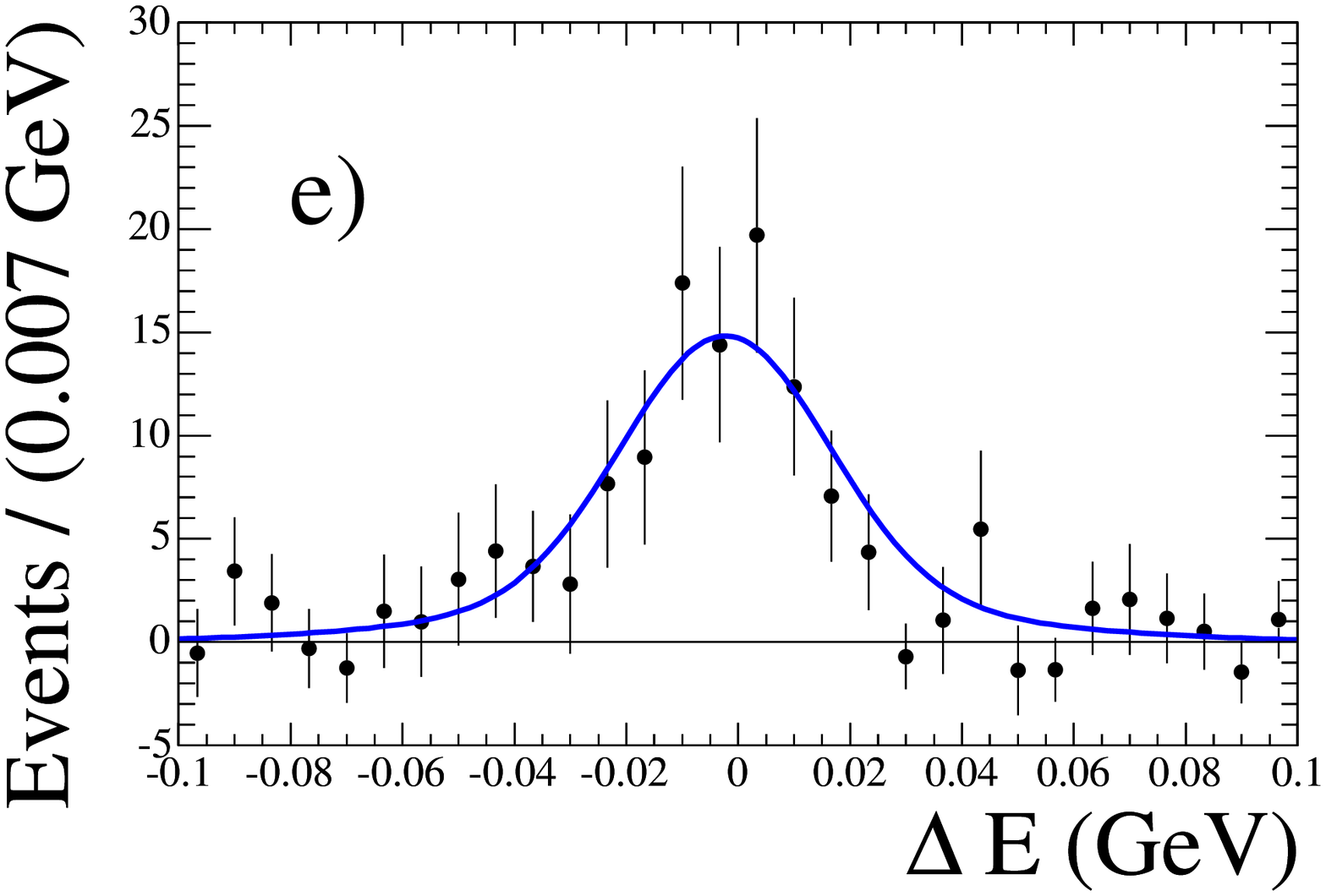} \\ 
\includegraphics{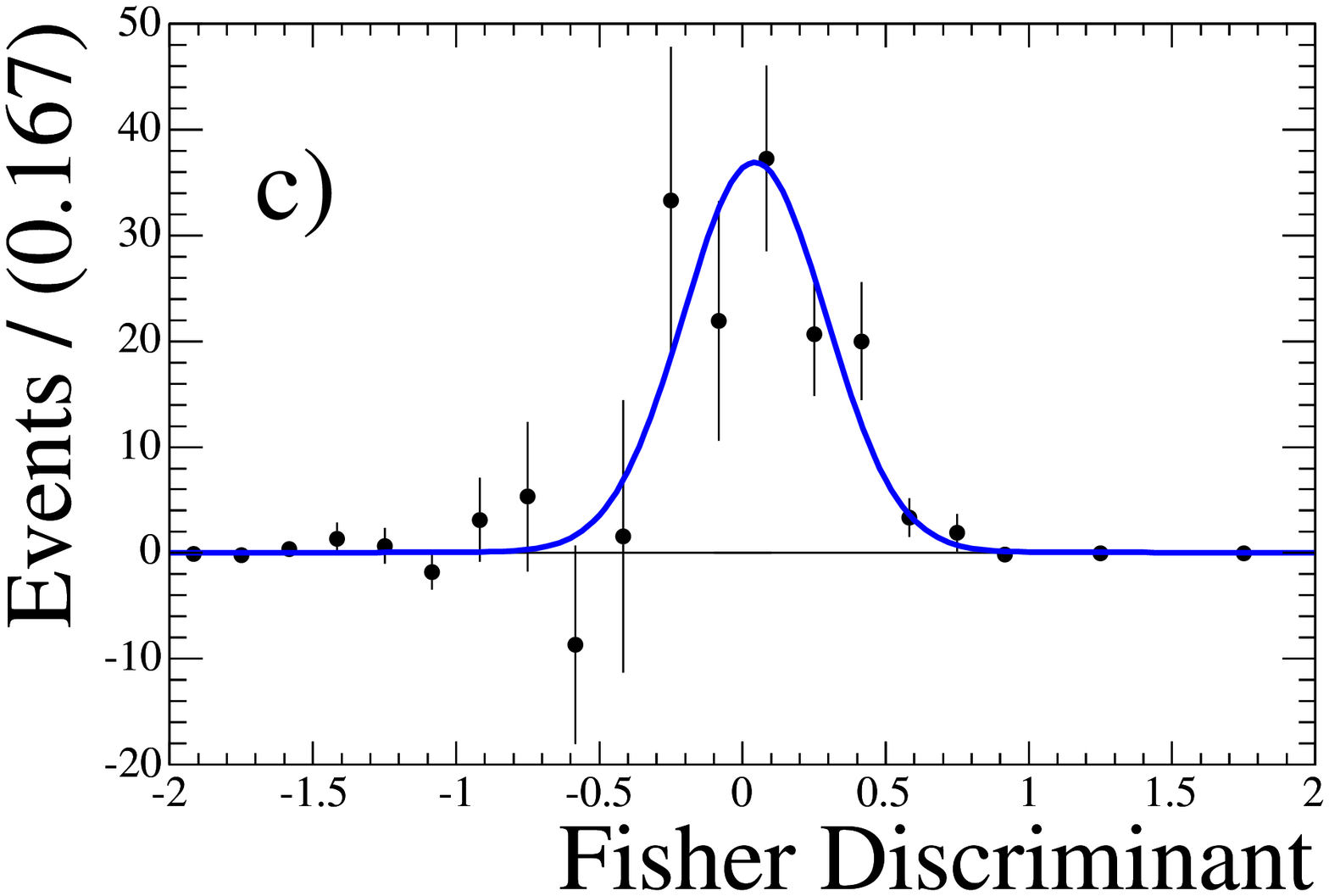} & 
\includegraphics{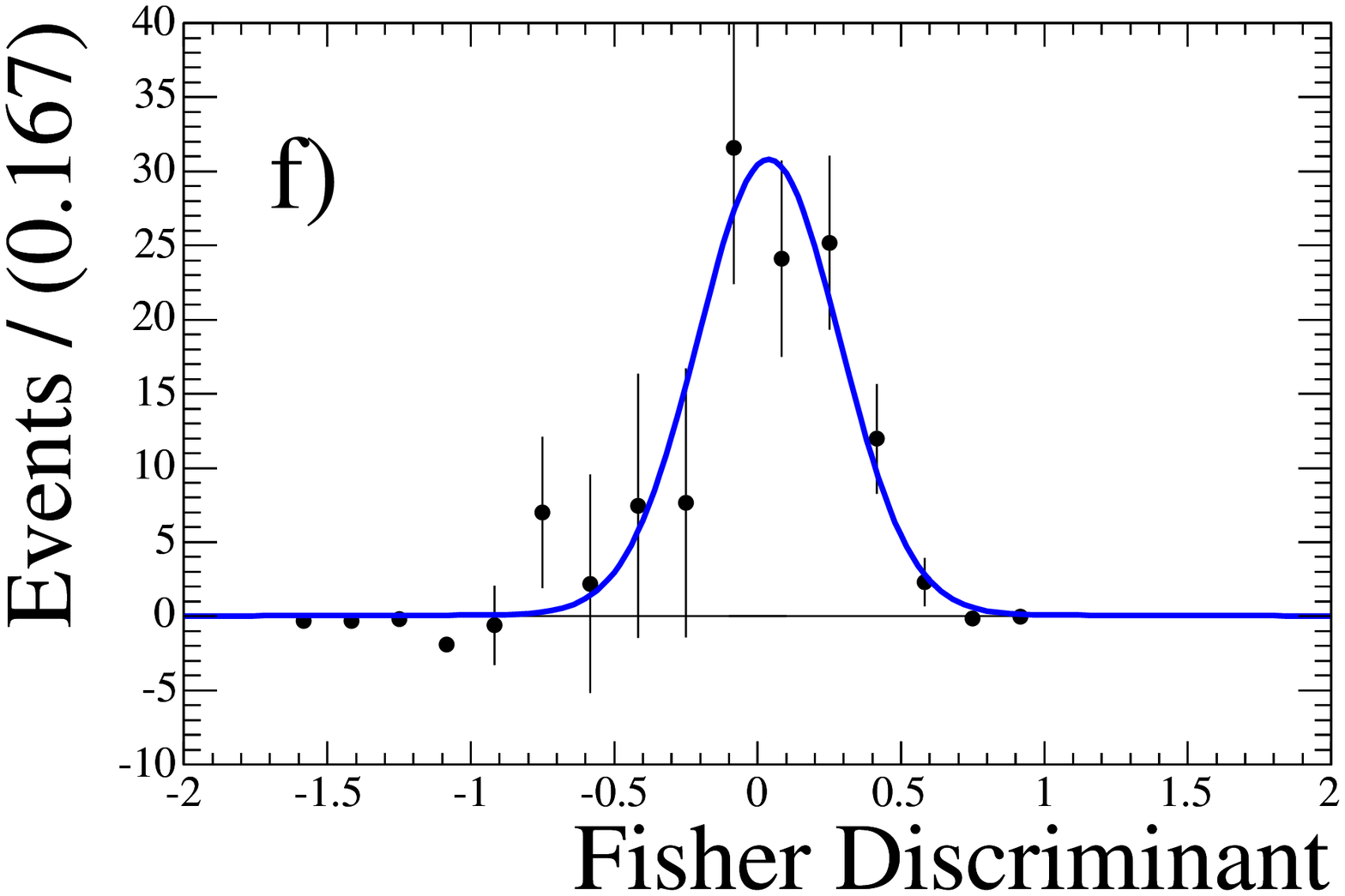}  
\end{tabular}
}
\caption{Maximum likelihood fits for signal distributions.  For 
$\Bz \to K^{*+}$$\pi^-$ the plots show (a) \mes, (b) \de, and
(c) the Fisher discriminant.  The (black) circles are the signal 
distribution extracted from the data with the method of Ref.~\cite{splot}
and the solid curve is the signal PDF that resulted from the maximum 
likelihood fit.  For $\Bz \to f_0 K^0$, plots show the distributions 
for (d) \mes, (e) \de, and (f) the Fisher discrminant, in an 
analogous fashion.
}\label{fig:fitproj2}
\end{figure}

\begin{figure}[!h]
\resizebox{0.6\columnwidth}{!}{
\begin{tabular}{c}
\includegraphics{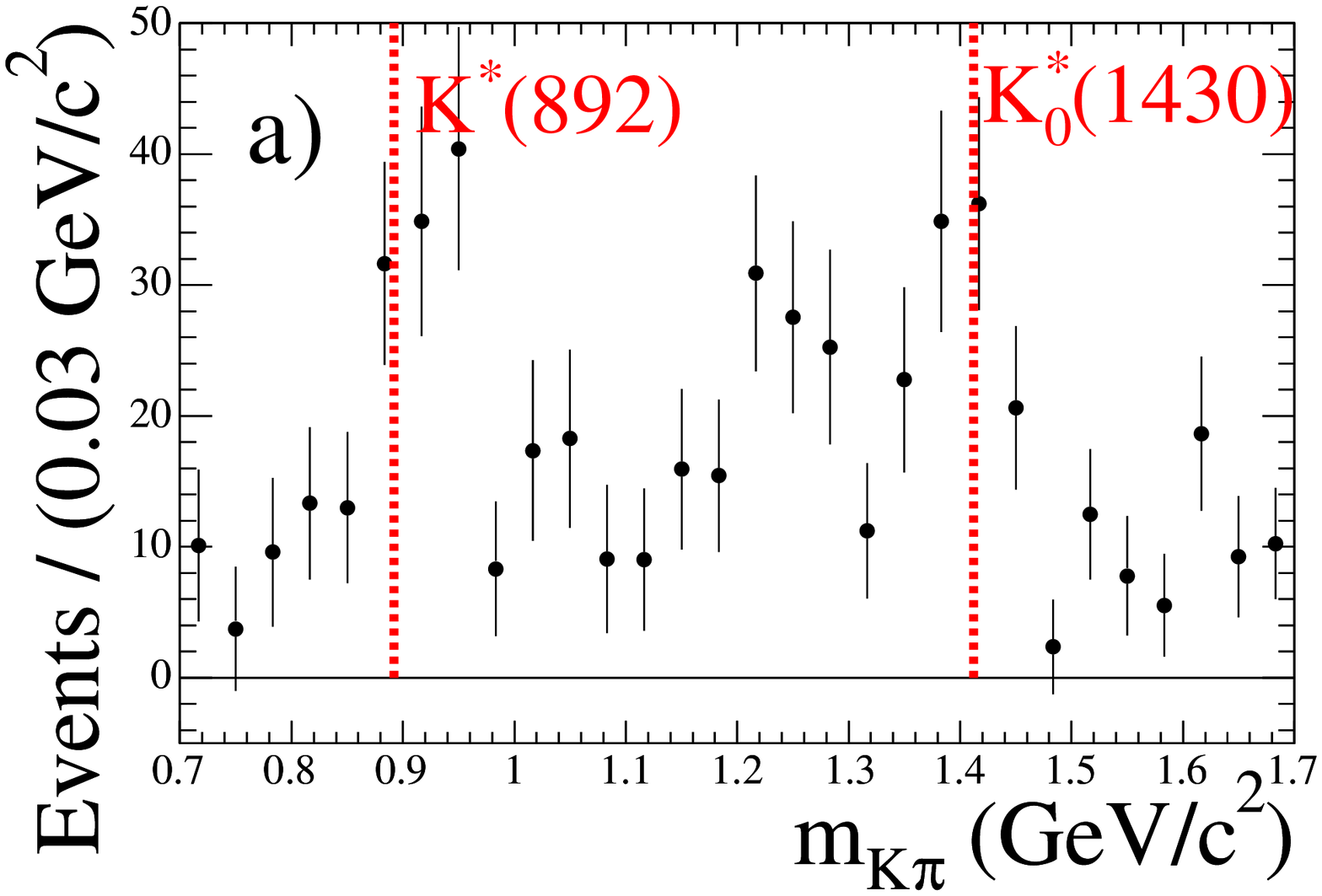}\\ \includegraphics{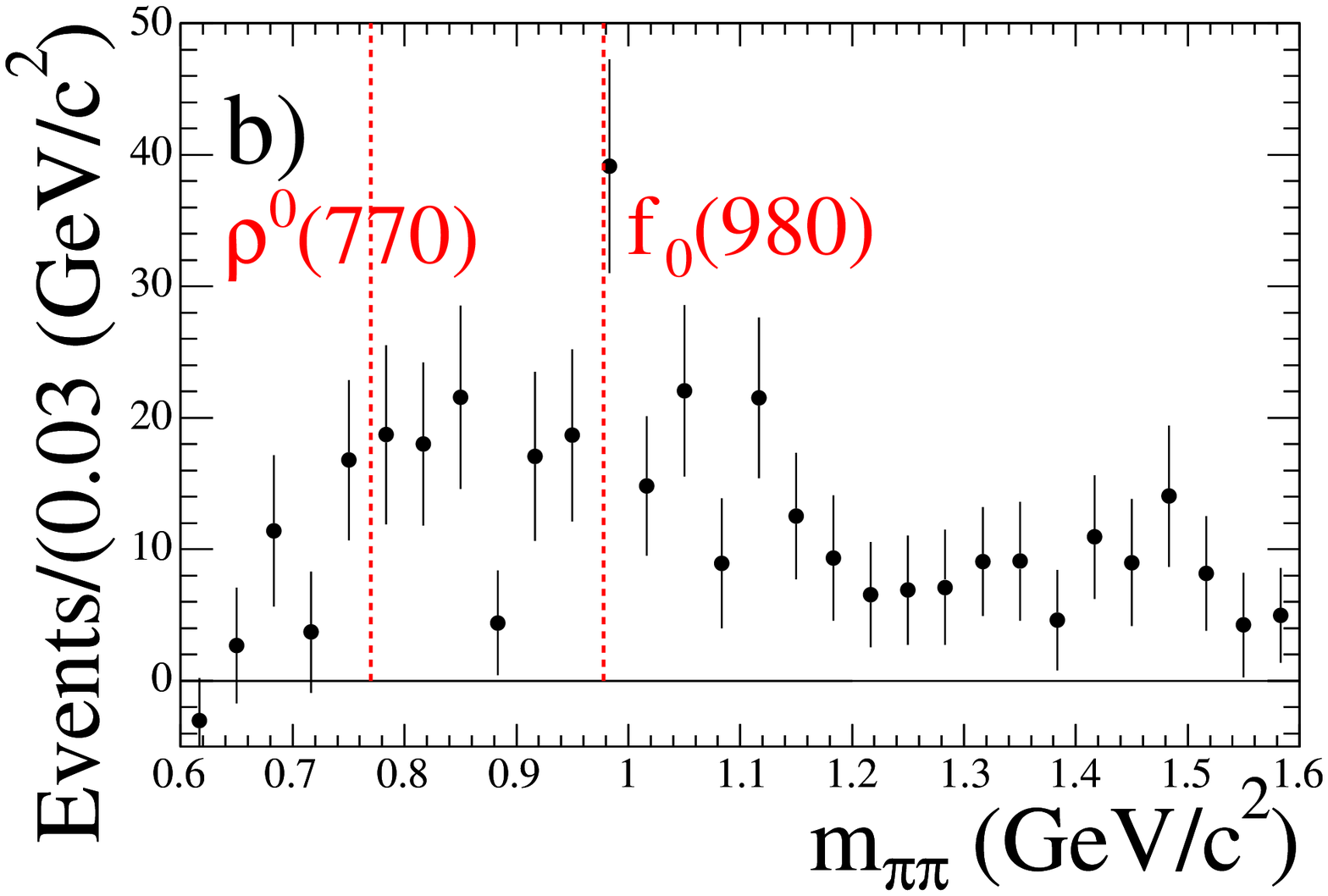} 
\end{tabular}
}
\caption{
a) shows the $m_{\KS\pi}$ signal distribution of $B^0 \to K^0\pi^+\pi^-$ 
candidates~\cite{splot}. The one-dimensional distribution is obtained by 
merging $m^2_{K\pi^+}$ and $m^2_{K\pi^-}$ into one ($m^2_{K\pi}$) by folding 
the Dalitz plane along the line corresponding to $m^2_{K\pi^+}$ = 
$m^2_{K\pi^-}$ in order to obtain the above $m_{K\pi}$ mass distribution.  
b) shows the $m_{\pip\pim}$ signal distribution of $B^0 \to K^0\pi^+\pi^-$ 
candidates~\cite{splot}.  The dashed lines indicate the expected mass 
of the labeled resonances.  }
\label{fig:projplots}  
\end{figure}

\begin{figure}[!h]
\resizebox{0.6\columnwidth}{!}{
\begin{tabular}{c}
\includegraphics{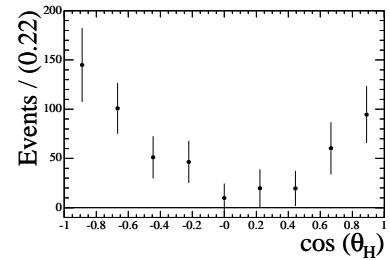}
\end{tabular}
}
\caption{
Distribution of the efficiency corrected cosine of the helicity angle, $\theta_H$, for
$\Bz \to K^{*\pm}\pi^{\mp}$ signal events. }
\label{fig:helplot}  
\end{figure}

Figure~\ref{fig:fitproj} shows the signal distributions for 
$B^0 \to K^0\pi^+\pi^-$ candidates and the distributions of events for all hypotheses.
Figure~\ref{fig:fitproj2} shows the signal distributions for both the 
$B^0$ $\rightarrow$ $K^{*+}$$\pi^-$ and $\Bz \to f_0 K^0$ channels.
The fitted signal yield and measured branching fraction are shown in 
Table \ref{tab:results} for all the modes under study. The 
average efficiency for $B^0 \to \KS\pi^+\pi^-$ signal events is 16.8\% and the continuum 
background yield is 79000 $\pm$ 280 events.
Figure~\ref{fig:projplots} shows the signal mass projections of $m_{\Ks\pi}$ and $m_{\pip\pim}$
using $B^0 \to K^0\pi^+\pi^-$ candidates. The $m_{\Ks\pi}$ distribution clearly shows a peak at 
0.9~GeV/$c^2$, corresponding to the $K^{*+}$(892) mass and there is a broad structure above 
1~GeV/$c^2$ that is the region where heavier kaon resonances can occur.
The $m_{\pip\pim}$ distribution shows evidence for resonance structure around 1 \gevcc that
corresponds to the $f_0$ and a broader structure below this that may be
attributed as the $\rho^0(770)$.  Figure~\ref{fig:helplot} shows the efficiency corrected signal
distribution of the cosine of the helicity angle, $\theta_H$, 
for $\Bz \to K^{*+}$$\pi^-$.
\begin{table}[!h]
\caption{Signal yields and branching fractions for $B^0 \to K^0\pi^+\pi^-$, $B^0$ 
$\rightarrow$ $K^{*+}\pi^-$ and $B^0$ $\rightarrow$ $f_0 K^0$ 
where the first uncertainty is statistical and where, in the case of the branching fraction 
measurements, the second uncertainty is systematic and any third uncertainty is due to
possible interference effects. The efficiency of selecting $B^0$ 
$\rightarrow$ $K^{*+}(\to \KS\pip) \pi^-$ and $B^0$ $\rightarrow$ $f_0(\to\pip\pim) \KS$ 
events was found to be 24\% and 27\% respectively, whilst the continuum background
yields were 7300 $\pm$ 86 events and 13000 $\pm$ 110 events respectively. The $B^0$ $\rightarrow$ 
$K^{*+}\pi^-$ branching fraction takes into account that ${\cal B}(K^{*+} \to K^0\pi^+) 
= 2/3$, assuming isospin symmetry.}\label{tab:results}
\begin{tabular}{lcc}
\hline
Mode&
Signal Events&
Branching Fraction\\
&
Yield&
($\times$ 10$^{-6}$)\\
\hline
$B^0 \to K^0\pi^+\pi^-$ &
860 $\pm$ 47 &
43.0 $\pm$ 2.3 $\pm$ 2.3 \\
$B^0$ $\rightarrow$ $f_0 (\to\pip\pim)K^0$ &
120 $\pm$ 16&
5.5 $\pm$ 0.7 $\pm$ 0.6 $\pm 0.3$ \\
$B^0$ $\rightarrow$ $K^{*+}\pi^-$ &
140 $\pm$ 19 &
11.0 $\pm$ 1.5 $\pm$ 0.5 $\pm 0.4$ \\
\hline
\end{tabular}
\end{table}

Table~\ref{tab:sys} shows the systematic uncertainties that are
assigned to the branching fraction measurements. Control
channels in data and MC are used to assign uncertainties due to pion 
tracking, particle identification, and $\Ks$ reconstruction efficiency.
To calculate uncertainties due to the fitting procedure, a 
large number of MC samples are generated from the fitted PDFs, containing the 
amounts of signal and continuum events that are measured in data
and the number of $B$-background events that were anticipated for the
data set, as explained above. The 
differences between the generated and fitted values using these samples 
are used to ascertain the sizes of any biases. Small biases of the order 
of a few percent are observed that are a consequence of small correlations 
between fit variables and are therefore assigned as systematic uncertainties. 

The uncertainty of the $B$-background contribution to the fit is estimated by 
varying the measured branching fractions within their uncertainties. Each background 
is varied by $\pm 1\sigma$~\cite{pdg} and the effect on the fitted signal yield is added 
as a contribution to the uncertainty. For $B^0$ $\rightarrow$ $K^{*+}\pi^-$ 
there is an additional uncertainty in the $B$-background contributions due to the 
possible lineshapes of the $K_0^{*\pm}$(1430), which can alter the
amount of $B$ background expected.  
In order to assign a systematic uncertainty, fits to data are performed
using two parameterizations, a relativistic Breit--Wigner lineshape
and the LASS parameterization~\cite{LASS}.  The latter is
a coherent sum of a relativistic Breit--Wigner and an effective range term,
and is used in the analysis of $B^{\pm}\to K^{\pm}\pi^{\mp}\pi^{\pm}$~\cite{kpipi}.
The uncertainty 
due to simulated PDFs is obtained from the channel $B^0$ $\rightarrow$ 
$D^-(\rightarrow \Ks\pi^-)\pi^+$ and by varying the PDFs according to the 
precision of the parameters obtained from MC. In order to take correlations 
between parameters into account, the full correlation matrix is used when 
varying parameters. All PDF parameters that are originally fixed in the fit 
are then varied in turn and each difference from the nominal fit is combined 
and taken as a systematic uncertainty. The uncertainty in the efficiency is due to 
limited MC statistics, where over 1,000,000 MC events are generated for the 
decay $B^0 \to K^0\pi^+\pi^-$ and over 150,000 MC events 
are generated for the decays $B^0$ $\rightarrow$ $K^{*+}\pi^-$ and $B^0$ 
$\rightarrow$ $f_0\KS$. The same uncertainty in the number of \BB\
events is used for all channels. 
\begin{table}[!ht]
\caption{Summary of contributions to the systematic uncertainty in the
branching fractions measurements of $B^0 \to K^0\pi^+\pi^-$, 
$B^0$ $\rightarrow$ $K^{*+}\pi^-$ and $B^0$ $\rightarrow$ $f_0 K^0$. 
The uncertainties are shown as a percentage of the measured branching fraction.}\label{tab:sys}
\begin{tabular}{lccc}
\hline
Error &
$B^0\to K^0\pip\pim$ & 
$\Bz \to f_0 K^0 $ &
\kstarpi \\
source &
Error (\%)&
Error (\%)&
Error (\%) \\
\hline
\hline
Particle ID	&
1.9 &
1.9 &
1.9 \\ 
\hline
Tracking	&
1.6 &
1.6 &
1.6 \\ 
\hline
$\Ks$ efficiency	&
1.4		 &
1.6 &
1.5	\\
\hline
Fit Bias	&
1.7	 &
6.1 &	
2.6	\\
\hline
PDF params.	&
0.1	 &
0.1 &	
0.3	\\
\hline
$B$ background	&
4.2	 &
5.9 &	
2.0	\\
\hline
Efficiency	&
0.9	 &
0.1 &	
0.1	\\
\hline
No. of $B\overline{B}$	&
1.1 &
1.1 &
1.1 \\ 
\hline
\hline
\bf{TOTAL }	&
5.4	 &
9.1 &
4.5	\\
\hline
\hline
Interference 	&
- &
4.7 & 
4.0	\\
\hline
\end{tabular}
\end{table}

For the quasi two-body modes, possible interference effects between the 
final state modes were investigated by simulating the Dalitz plot using
the measured branching fractions and random phases.  The root-mean-squared
of the distribution of the branching fraction is taken to be the uncertainty.  

We measure the \CP-violating charge asymmetry for the decay $B^0$ $\rightarrow$ 
$K^{*+}$$\pi^-$ to be ${\cal A}_{K^*\pi} = 
-0.11 \pm 0.14 \pm 0.05$, where the first uncertainty is statistical 
and the second uncertainty is systematic. The charge asymmetry in
the background is expected to be zero, as is the charge asymmetry in signal 
and background of the self-tagging decay \dpi.  As a cross-check, these 
are measured to be $-$0.018 $\pm$ 0.009, $-$0.013 $\pm$ 0.029 and 0.005 
$\pm$ 0.031 respectively, where the uncertainties are statistical only.

The systematic uncertainty on ${\cal A}_{K^*\pi}$ is calculated by considering 
contributions due to track finding, particle identification, fit biases 
and $B$-background asymmetry uncertainties. Biases due to track finding 
and particle identification were found to be negligible. The fit-bias 
contribution to the systematic uncertainty is calculated using a large number 
of MC samples. The contribution from $B$ background is calculated by 
varying the number of expected events within their uncertainties~\cite{pdg} and by assuming a 
conservative \CP-violating asymmetry of $\pm$0.5 as there are no 
available measurements for these decays. The resulting systematic 
uncertainty on the asymmetry is measured to be $\pm0.05$.

In summary, the branching fractions for $B^0 \to K^0\pi^+\pi^-$, 
$B^0$ $\rightarrow$ $K^{*+}\pi^-$, and $\Bz \to f_0 (\to \pip\pim)
K^0$ decaying to a $\Ks\pi^+\pi^-$ 
state are measured and all agree with previous 
measurements~\cite{cleo,belle,f0Ks,kspipi}. We measure the 
direct \CP-violating parameter ${\cal A}_{K^*\pi}$ for the decay $B^0$ 
$\rightarrow$ $K^{*+}\pi^-$, with no evidence of \CP\ violation 
with the statistics used.  These results supersede the previous
results of the \babar\ Collaboration~\cite{f0Ks,kspipi}.

\input acknow_PRL.tex

\end{document}

%% file: symbols.tex
\newcommand{\beq}{\begin{equation}}
\newcommand{\eeq}{\end{equation}}
\newcommand{\bea}{\begin{eqnarray}}
\newcommand{\eea}{\end{eqnarray}}

\newcommand{\DE}{\ensuremath{\Delta E}}
\newcommand{\de}{\ensuremath{\Delta E}}

\newcommand{\pvec}{{\bf p}}

\def\Y#1S{{\Upsilon\rm(#1S)}}
\def\Ks{{K^0_{\scriptscriptstyle S}}}

\def\ra{\rightarrow}

%% file: acknow_PRL.tex
We are grateful for the excellent luminosity and machine conditions
provided by our \pep2\ colleagues, 
and for the substantial dedicated effort from
the computing organizations that support \babar.
The collaborating institutions wish to thank 
SLAC for its support and kind hospitality. 
This work is supported by
DOE
and NSF (USA),
NSERC (Canada),
IHEP (China),
CEA and
CNRS-IN2P3
(France),
BMBF and DFG
(Germany),
INFN (Italy),
FOM (The Netherlands),
NFR (Norway),
MIST (Russia), and
PPARC (United Kingdom). 
Individuals have received support from CONACyT (Mexico), A.~P.~Sloan Foundation, 
Research Corporation,
and Alexander von Humboldt Foundation.